\documentclass[twocolumn,english,prl,twocolumn,showpacs,superscriptaddress]{revtex4-1}
\usepackage[T1]{fontenc}
\usepackage[latin9]{inputenc}
\usepackage{babel}
\usepackage{amssymb}
\usepackage{graphicx}
\usepackage{esint}
\usepackage[unicode=true,pdfusetitle,
 bookmarks=true,bookmarksnumbered=false,bookmarksopen=false,
 breaklinks=false,pdfborder={0 0 1},backref=false,colorlinks=false]
 {hyperref}

\makeatletter
\@ifundefined{textcolor}{}
{%
 \definecolor{BLACK}{gray}{0}
 \definecolor{WHITE}{gray}{1}
 \definecolor{RED}{rgb}{1,0,0}
 \definecolor{GREEN}{rgb}{0,1,0}
 \definecolor{BLUE}{rgb}{0,0,1}
 \definecolor{CYAN}{cmyk}{1,0,0,0}
 \definecolor{MAGENTA}{cmyk}{0,1,0,0}
 \definecolor{YELLOW}{cmyk}{0,0,1,0}
 }

\usepackage{hyperref}

\makeatother

\begin{document}

\title{Quantum simulation of expanding space-time with tunnel-coupled condensates}

\author{Clemens Neuenhahn and Florian Marquardt}

\affiliation{Friedrich-Alexander-Universität Erlangen-Nürnberg,\\
Institute for Theoretical Physics II, Staudtstr.7, 91058 Erlangen,
Germany}
\begin{abstract}
We consider two weakly interacting quasi-1D condensates of cold bosonic
atoms. It turns out that a time-dependent variation of the tunnel-coupling
between those condensates is equivalent with the spatial expansion
of a one-dimensional toy-Universe with regard to the dynamics of the
relative phase field. The dynamics of this field is governed by the
quantum sine-Gordon equation. Thus, this analogy could be used to
'quantum simulate' the dynamics of a scalar, interacting quantum field
on an expanding background. We discuss, how to observe the freezing
out of quantum fluctuations during an accelerating expansion in a
possible experiment. We also discuss an experimental protocol to study
the formation of sine-Gordon breathers in the relative phase field
out of quantum fluctuations. 
\end{abstract}
\maketitle
The recent progress in coherently controlling systems of cold atoms
(e.g.,$\,$ \cite{2002_GReinerCollapseRevival,2006_Kinoshita,2007_Hofferbeth,2011_ZollerSimulation}
, see also \cite{2008_Bloch_ColdGases} and references therein), stimulated
a lot of research concerned with employing these experimental systems
to 'quantum simulate' prototypical quantum many-body models (see e.g.,
\cite{2012_JordanPhi4,2009_Nori}). Particularly fascinating are ideas
concerned with simulating quantum many-body physics on curved space-time
('analog gravity') (see, e.g., \cite{2000_ZollerBlackHole,2004_Schulz,2007_Schuetzhold,2007_Jain,2010_Prain,1981_Unruh,2005_weinfurtner}
and \cite{2011_AnalogGravity}, as well as references therein) connecting
concepts and techniques from cosmology and condensed matter (see \cite{2005_Schuetzhold}). 

\begin{figure}
\includegraphics[width=1\columnwidth]{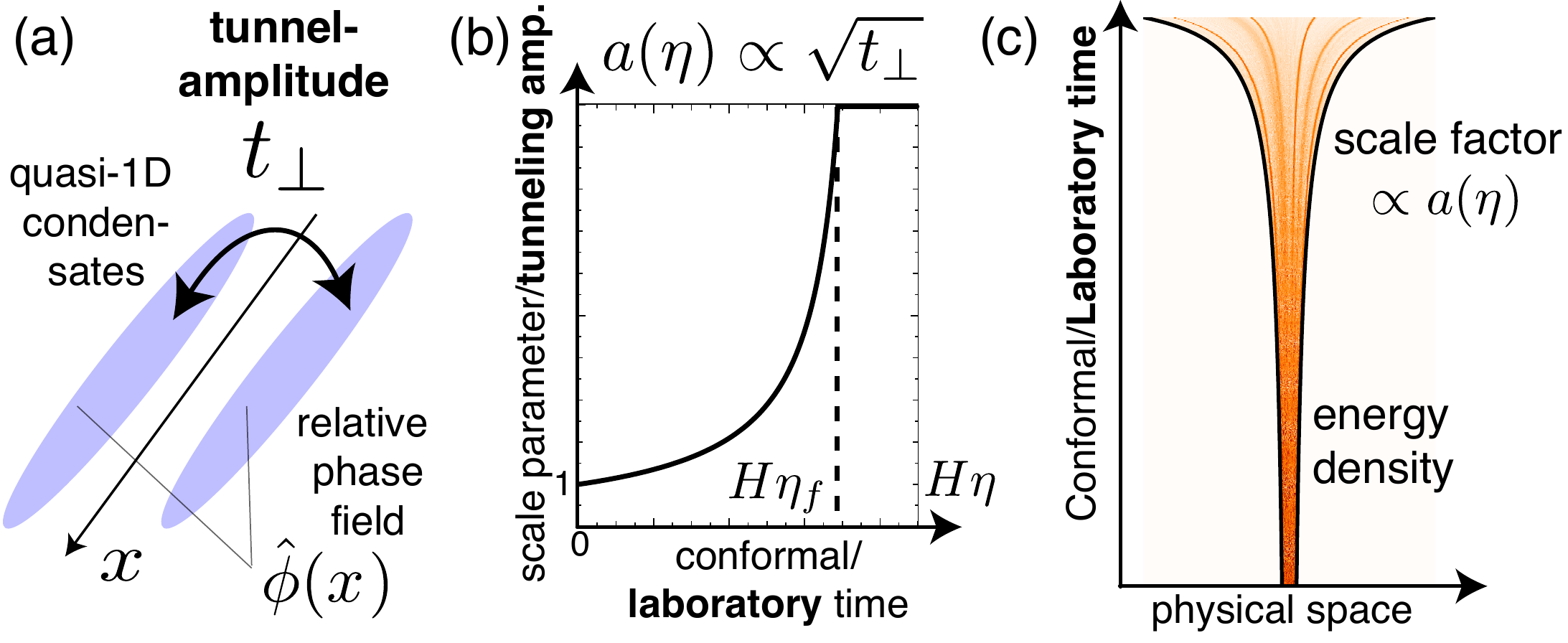}

\caption{(Color online) Quantum simulator. (a) Sketch of the quantum simulator
consisting of a pair of tunnel-coupled 1D bosonic condensates. The
tunnel amplitude $t_{\perp}$ is modulated time-dependently. It can
be related to the scale parameter $a$ (b) describing the (exponential)
expansion of a 1+1 dimensional 'Universe'. The laboratory time (condensate
coordinate $x$) is identified with the conformal time $\eta$ (co-moving
coordinate) (see main text). The relative phase field between the
condensates $\hat{\phi}$ represents the scalar quantum field to be
simulated. (c) Semiclassical evaluation of the physical energy density
of the simulated quantum field (here displayed in physical coordinates
$x_{{\rm ph}}=ax$). One can observe the structure formation out of
zero-point quantum fluctuations during the expansion (see main text).
\label{Fig_Cosmo_Pub_1}}
 
\end{figure}
Here, it is argued that a pair of tunnel-coupled, quasi-1D, bosonic
condensates (as realized in \cite{2005_SchmiedmayerMAI} and related
experiments) can be employed for simulating an interacting, scalar
quantum field on top of an expanding 1+1 dimensional space-time (Fig.$\,$\ref{Fig_Cosmo_Pub_1}a,c).
This scalar field is represented by the relative phase field between
the condensates. As argued in \cite{2007_AnatoliSG,2012_Neuenhahn},
at low energies, its dynamics is described by the quantum sine-Gordon
model. It turns out that in this setup, one can simulate the expansion
of the 1D toy-Universe simply by varying the tunnel-amplitude according
to a suitable protocol (Fig.$\,$\ref{Fig_Cosmo_Pub_1}b). In the
experiments (e.g., \cite{2005_SchmiedmayerMAI}) the tunnel-amplitude
itself can be largely tuned and the field dynamics can be directly
visualized by means of matter-wave interferometry. The 'quantumness'
of the relative phase-field dynamics depends on the interaction strength
within each condensate (which is easily tunable, e.g., by adapting
the 1D condensate density). As far as we are dealing with non-linear
dynamics, we consider the weakly interacting, 'semiclassical' limit
(e.g., \cite{2010_AnatoliTWA,2012_Neuenhahn,1986_FowlerPRL}) and
employ the truncated Wigner approximation (TWA) (e.g., \cite{2010_AnatoliTWA}).
In contrast, the proposed quantum simulator would allow to explore
also the deep quantum regime - in fact, this is its main purpose. 

In the following, we start with deriving an effective Hamiltonian
description of an interacting quantum field on an expanding 1+1 dimensional
space time. In a second step, we introduce the experimental 'quantum
simulator' - the tunnel-coupled condensates - and eventually discuss
two of the possible effects, which could be explored. 

The first effect is the well known 'freezing' of quantum fluctuations
and the related 'cosmological particle production' during an accelerating
expansion. This purely linear, though most fundamental effect is made
responsible for the structure formation in the very early universe
\cite{2005_Mukhanov}. In the present experiments, one can only perform
single measurements of the relative phase field per run. Fortunately,
this mode freezing also manifests itself in the spatial fluctuation
spectrum of the field and thus, in principle, is detectable with single
measurements per run. Although, strictly speaking, there is no need
for a quantum simulation of the exactly solvable linear dynamics,
observing the freeze-out of quantum fluctuations in the experiment
would nevertheless be exciting (see for instance \cite{2007_Schuetzhold,2004_Schulz})
and constitute an important check on the setup. 

The second feature is the generation of localized, macroscopic structures
during the expansion out of quantum fluctuations. This pattern formation
involves the full non-linearity of the underlying sine-Gordon field
theory and was also observed in the static case \cite{2012_Neuenhahn}.
In contrast to \cite{2012_Neuenhahn}, for an exponentially fast expansion
this happens only for small enough expansion rates. At large expansion
times, these patterns seem to turn into standing sine-Gordon breathers
simply drifting apart from each other. We argue how to detect signatures
of this 'Hubble' drift experimentally. In cosmology, e.g., dealing
with the preheating following inflation, excitations like this (e.g.,
in the scalar inflaton field) are sometimes denoted as 'oscillons'
(e.g., \cite{2010_Amin3D,2010_Amin,2009_Gleiser,2003_Gleiser,2010_Hertzberg,2012_AminPRL}).
A full experimental quantum simulation would allow investigating the
formation and persistence of these excitations on an expanding background,
even in regimes where quantum effects become very important for the
dynamics. 

\emph{Quantum field on curved space-time.} - The space-time action
of a classical, scalar field theory in 1+1 dimensions is given by
\cite{2005_Mukhanov} ($c=1$, $\hbar=1$) 
\begin{eqnarray}
\mathcal{S} & = & \frac{1}{2}\int dx^{2}\sqrt{\left|{\rm det}g_{\mu\nu}\right|}\left[(\partial_{\mu}\chi)g^{\mu\nu}(\partial_{\nu}\chi)-2V(\chi)\right],\label{Action}
\end{eqnarray}
where $g_{\mu\nu}$ denotes the metric (with $g_{\mu\nu}g^{\nu\gamma}=\delta_{\mu}^{\gamma}$),
$\partial_{\mu}\chi=\partial\chi/\partial x^{\mu}$ and $V(\chi)$
is an arbitrary potential. We are interested in an homogeneous, spatially
expanding space-time described by the Friedmann-Robertson-Walker metric
(FRW) (see for instance \cite{2005_Mukhanov}, $ds^{2}=g_{\mu\nu}dx^{\mu}dx^{\nu}$)
\begin{eqnarray}
ds^{2} & = & d\tau^{2}-a^{2}(\tau)dx^{2}.\label{metric}
\end{eqnarray}
Here, $x$ denotes co-moving coordinates which are related to physical
coordinates $x_{{\rm ph}}$ via the scale parameter $a(\tau)$ as
$x_{{\rm ph}}=ax$. The time $\tau$ denotes the cosmological time,
i.e., the proper time of a co-moving observer. In reality, the dynamics
of $a$ is determined by Einstein's equations. Here, we treat $a(\tau)$
as a given function of time, which will be specified below. 

In 1+1 dimensions, the Lagrangian corresponding to the action Eq.$\,$(\ref{Action})
takes an intriguingly simple form using the so-called conformal time
$\eta$ with $d\eta=d\tau/a$: 
\begin{eqnarray}
L & = & \frac{1}{2}\int dx\,\left[\left(\partial_{\eta}\chi\right)^{2}-\left(\partial_{x}\chi\right)^{2}-2a^{2}(\eta)V(\chi)\right].\label{Lagrangian}
\end{eqnarray}
Note that light geodesics ($ds^{2}=0$) are described by $x(\eta)=\pm\eta+{\rm const}$.
The quantization of the field theory follows the standard prescription
(e.g., \cite{2005_Mukhanov}). First, from Eq.$\,$(\ref{Lagrangian}),
we construct the canonically conjugated field $\Pi_{\chi}(x,\eta)\equiv\frac{\delta\mathcal{\mathcal{L}}}{\delta\partial_{\eta}\chi}=\partial_{\eta}\chi$.
In a second step, we promote the fields to operators demanding that
$[\hat{\chi}(x,\eta),\hat{\Pi}_{\chi}(x',\eta)]=i\delta(x-x')$. Eventually,
we can switch to the Hamiltonian formulation introducing the time-dependent
Hamiltonian 
\begin{eqnarray}
\hat{H}(\eta) & = & \frac{1}{2}\int dx\,\left[\hat{\Pi}_{\chi}^{2}+\left(\partial_{x}\hat{\chi}\right)^{2}+2a^{2}(\eta)V(\hat{\chi})\right].\label{Hamiltonian_Effective_Cosmo}
\end{eqnarray}
Note that all effects of the expanding space-time are now encoded
in the time-dependence of $\hat{H}(\eta)$ (cf. \cite{2005_Schuetzhold,2012_SchuetzholdUnruh,2010_Amin1D}).

\emph{Tunnel coupled condensates as quantum simulator.} \emph{-} Here,
we propose a quantum simulation of the field $\hat{\chi}(x,\eta)$
for the special case of a sine-Gordon potential $V=-m_{0}^{2}\beta^{-2}\cos(\beta\hat{\chi})$.
This potential has several interesting properties: The corresponding
field theory is interacting and integrable. Second, the sine-Gordon
potential appears in the so-called 'natural inflation' scenario \cite{2007_Linde}.
Third, the sine-Gordon potential supports the formation of 'quasibreathers'
\cite{2012_Neuenhahn} (in the cosmology literature denoted as 'oscillons',
e.g., \cite{2009_Gleiser,2008_Farhi,2012_AminPRL,2003_Gleiser} and
references therein).

The quantum simulator (Fig.$\,$$\,$\ref{Fig_Cosmo_Pub_1}a) consists
of two tunnel-coupled quasi-1D condensates of cold, bosonic atoms
(e.g., \cite{2005_SchmiedmayerMAI}) with a time-dependent tunnel
amplitude $t_{\perp}$. The laboratory time is identified with the
conformal time $\eta$. At low energies, the dynamics of the relative
phase field $\beta\hat{\phi}(x,t)/\sqrt{2}=(\hat{\phi}_{1}-\hat{\phi}_{2})/\sqrt{2}$
can be described by the quantum sine-Gordon model \cite{2007_AnatoliSG}
(the sound velocity $v_{s}=1$)
\begin{eqnarray}
\hat{H}_{{\rm SG}} & = & \frac{1}{2}\int_{0}^{L}dx\,\left[\hat{\Pi}^{2}+\left(\partial_{x}\hat{\phi}\right)^{2}-\frac{2m^{2}(\eta)}{\beta^{2}}\cos\beta\hat{\phi}\right].\label{H_SG}
\end{eqnarray}
The relative phase field and the relative density variations $\sqrt{2}\beta^{-1}\hat{\Pi}\equiv\frac{\hat{\pi}_{1}-\hat{\pi}_{2}}{\sqrt{2}}$
form a canonical pair. The tunnel amplitude enters the mass term $m^{2}(\eta)=2\beta^{2}\rho_{0}t_{\perp}(\eta)$,
where $\rho_{0}$ is the mean density per condensate. The Luttinger
liquid description should be reliable as long as the typical lengthscale
of Eq.$\,$(\ref{H_SG}), set by $\sqrt{\hbar v_{s}}/m$, is much
larger than the healing length of the condensates $\xi_{h}$ \cite{2012_Neuenhahn,2007_AnatoliSG}.
This can always achieved by choosing a sufficiently small tunnel amplitude
$t_{\perp}(\eta)$. The parameter $\beta$ is related to the Luttinger
parameter $K$ as $\beta=\sqrt{2\pi/K}$. For weak interactions $\beta\ll1$.
This is the limit we are considering here. It can be shown that $\beta$
plays the role of Plank's constant \cite{2010_AnatoliTWA} and $\beta\ll1$
corresponds to the semiclassical limit of the quantum sine-Gordon
model (see also, e.g., \cite{1986_Fowler}). The analysis here (as
far as we are dealing with non-linear dynamics) is based on the semiclassical
TWA. However, in the experiment one can go deep into the quantum regime
corresponding to larger $\beta$ (a rather broad range of values up
to $K\sim50$ is realizable, e.g. \cite{2010_KruegerPRL}). The following
identifications connect the quantum simulator and the quantum field
theory on an expanding background. Identifying the fields $\hat{\chi}\leftrightarrow\hat{\phi}$
and $m^{2}(\eta)=m_{0}^{2}a^{2}(\eta)$ {[}and thus $a(\eta)\leftrightarrow\sqrt{t_{\perp}(\eta)/t_{\perp}(0)}]$,
the dynamics of the relative phase field simulates $\hat{\chi}$ in
conformal time and co-moving coordinates. In the remainder, we will
always argue in terms of the field $\hat{\phi}$, i.e., we analyze
Eq.$\,$(\ref{H_SG}). 

\emph{Scale parameter and initial state.} - We consider an exponential
expansion $a(\tau)=e^{H\tau}$ with the 'Hubble constant' $H=\dot{a}(\tau)/a(\tau)$.
Choosing $a(0)=1$, the conformal\textbf{ }time is given by $\eta=H^{-1}\left[1-e^{-H\tau}\right]$
and correspondingly $a(\eta)=[1-H\eta]^{-1}$ (see Fig.$\,$\ref{Fig_Cosmo_Pub_1}b)
with $\eta\in[0,\eta_{f}]$ and $\eta_{f}<1/H$. At $\eta_{f}$, the
expansion ends and $a(\eta>\eta_{f})=a_{f}$. The dimensionless parameter
$H/m_{0}$ compares the expansion time-scale and the typical internal
time scale (at short times) of the system and plays a crucial role
throughout the following. 

At $\eta=0$, we start in the ground state of massive phonons with
a small mass $m_{0}$. In particular, $m_{0}$ is chosen much smaller
than the UV cutoff $1\sim1/\xi_{h}$ (all TWA simulations are performed
on a lattice with lattice constant set to one, while keeping $m_{0}a(\eta)\ll1$
throughout the whole simulation). For finite system size and $\beta\ll1$,
the center-of-mass mode (COM mode) $\hat{\Phi}(\eta)\equiv\frac{1}{L}\int_{0}^{L}dx\,\hat{\phi}(x,\eta)$
needs some special attention. For $L/\xi_{h}\gg1$, we can treat the
COM mode classically such that $\hat{\phi}(x,\eta)\simeq\Phi(\eta)+\delta\hat{\phi}(x,\eta)$.
Before, the expansion starts, the COM mode is tuned to some value
$\Phi(0)\simeq\Phi_{0}$ with $\beta\Phi_{0}\in[0,\pi]$ ($\Phi'(0)/m_{0}\simeq0)$.
As argued in \cite{2012_Neuenhahn}, such an initial state can be
achieved by slowly splitting a single condensate followed by applying
a potential gradient between the condensates to tune $\Phi_{0}$.

\emph{Freezing out of quantum fluctuations. }- 
\begin{figure}
\includegraphics[width=1\columnwidth]{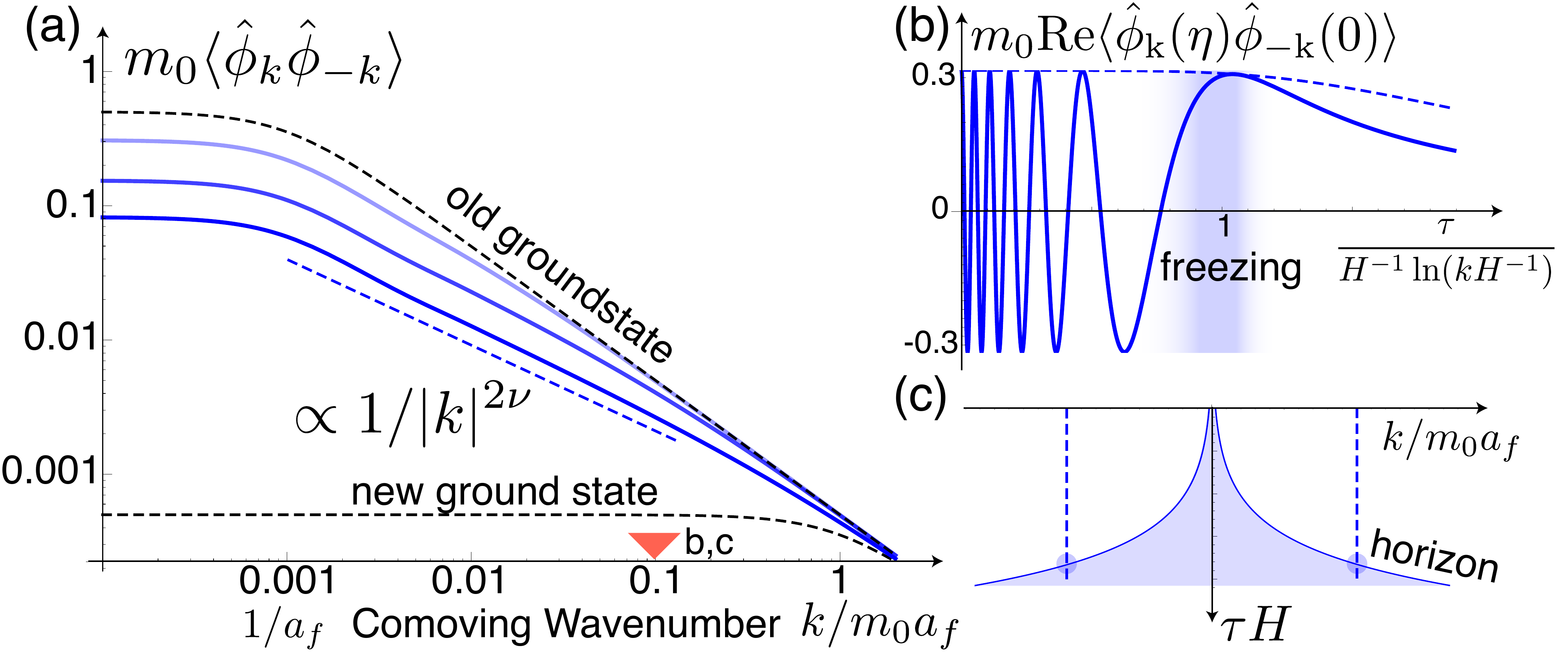}

\caption{(Color online) Mode 'freezing'. (a) Plot of the spectrum at $a=19.6,166.8,1000$
(blue lines, top to bottom). Here, $H/m_{0}=2.6$ and $m_{0}=0.05$.
One observes the spectral tilt (dashed, blue line) resulting from
the freezing out of modes. Dashed lines show the corresponding ground
state spectra for $a=1$ and $a=a_{f}$. (b) 'Freezing out' of modes
(here $k/a_{f}m_{0}=0.1$). While the suppression of the 1D field
fluctuations at later times is non-universal (e.g., it depends on
the spatial dimension), the mechanism of the expansion-induced mode
freezing is the same as in higher dimensions. (c) Sketch of the horizon
crossing \label{Fig_Cosmo_Pub_2}}
\end{figure}
One of the fascinating results of modern cosmology is that the structure
formation in the very early Universe seems to have been seeded by
quantum fluctuations \cite{2005_Mukhanov}. This result is truly amazing,
as on cosmological scales zero-point fluctuations are tiny. However,
it seems that an exponential expansion of the very early Universe
(inflationary stage) led to a 'freezing' of quantum fluctuations and
stretched them to cosmological scales. One can reformulate this basic
mechanism in a condensed matter language (e.g., \cite{2005_Schuetzhold}).
In this terminology, the inflationary expansion corresponds to a rapid,
non-adiabatic 'quench' (see, e.g., \cite{2008_Polkovnikov}) producing
a large number of excitations ('cosmological particle creation', cf.
\cite{2007_Schuetzhold}). 

A very similar effect should be observable in the considered 1+1 dimensional
toy-Universe. For this purpose, we consider the case $\Phi_{0}=0$.
For small enough $\beta$ and finite system size, one can safely expand
the cosine-potential to lowest order $m^{2}(\eta)\beta^{-2}(1-\cos\beta\hat{\phi})\approx\frac{m^{2}(\eta)}{2}\hat{\phi}^{2}$
yielding a theory of massive phonons (cf. \cite{2010_CazalillaNJP,2010_Grandi,2008_Grandi,2008_Polkovnikov,2005_weinfurtner})
on an expanding 1+1 FRW space-time. According to Eq.$\,$(\ref{Hamiltonian_Effective_Cosmo}),
the dynamics of the modes $\hat{\phi}_{k}$ {[}$\hat{\phi}(x,\eta)\equiv\hat{\Phi}+\frac{1}{\sqrt{L}}\sum_{k\neq0}e^{ikx}\hat{\phi}_{k}(\eta)${]}
follows as 
\begin{eqnarray}
\hat{\phi}_{k}''(\eta)+\left[k^{2}+m_{0}^{2}a^{2}(\eta)\right]\hat{\phi}_{k}(\eta) & \approx & 0.\label{EOMModes}
\end{eqnarray}
The general solution of Eq.$\,$(\ref{EOMModes}) is given in the
Supplement. For $|k|/aH\gg|\nu^{2}-\frac{1}{4}|$ (where $\nu=\frac{1}{2}\sqrt{1-4m_{0}^{2}/H^{2}}$),
the modes evolve freely. However, for $|k|/a(\eta)H\sim|\nu^{2}-\frac{1}{4}|$
(cf. Fig.$\,$\ref{Fig_Cosmo_Pub_2}b,c), the modes 'freeze'. Most
importantly, this 'freezing' also manifests in the spectrum $\langle\hat{\phi}_{k}(\eta)\hat{\phi}_{-k}(\eta)\rangle$
(cf. \cite{2004_Schulz,2005_Schuetzhold,2005_Mukhanov}). This, in
principle allows to observe this effect with a single measurement
of the relative phase field $\hat{\phi}(x,\eta)$ per run. For the
considered protocol (restricting to $\eta<\eta_{f}$), we obtain for
all modes with $|k|\gg H|\nu^{2}-\frac{1}{4}|,m_{0}$ (starting inside
the {}``horizon'') and $|k|/a(\eta)\ll H$ (see Fig.$\,$\ref{Fig_Cosmo_Pub_2}a)
\begin{eqnarray}
\langle\hat{\phi}_{k}(\eta)\hat{\phi}_{-k}(\eta)\rangle & \propto & \frac{1}{a^{1-2\nu}}\frac{H^{2\nu-1}}{|k|^{2\nu}}.\label{eq:-2-3}
\end{eqnarray}
Modes with $|k|/a(\eta)\gg H|\nu^{2}-\frac{1}{4}|$ remain close to
the initial ground state of massive phonons with $m_{0}$ yielding
$\langle\hat{\phi}_{k}(\eta)\hat{\phi}_{-k}(\eta)\rangle\simeq1/2|k|$.
Here, we consider a fast expansion with $H/m_{0}>2$. In the limit
$H/m_{0}\rightarrow\infty$, $\nu$ approaches $1/2$ corresponding
to a 'sudden quench', which leaves the spectrum unchanged. In a possible
experiment, one could detect the different power-laws by probing the
longitudinal (relative) phase coherence \cite{2006_Gritsev_InterfBosons}
of the condensates on different length-scales (very similar to \cite{2011_SchmiedmayerPrethermalization}).
It can be easily checked that the finite $a'(0)$ does not influence
the 'particle production' (indicated by the deviations from the instantaneous
ground state spectrum {[}see Eq.$\,$(\ref{eq:-2-3}){]}), which happens
happens during the evolution. 

\emph{Formation of breathers out of quantum fluctuations.} \textbf{-}
\begin{figure}
\includegraphics[width=1\columnwidth]{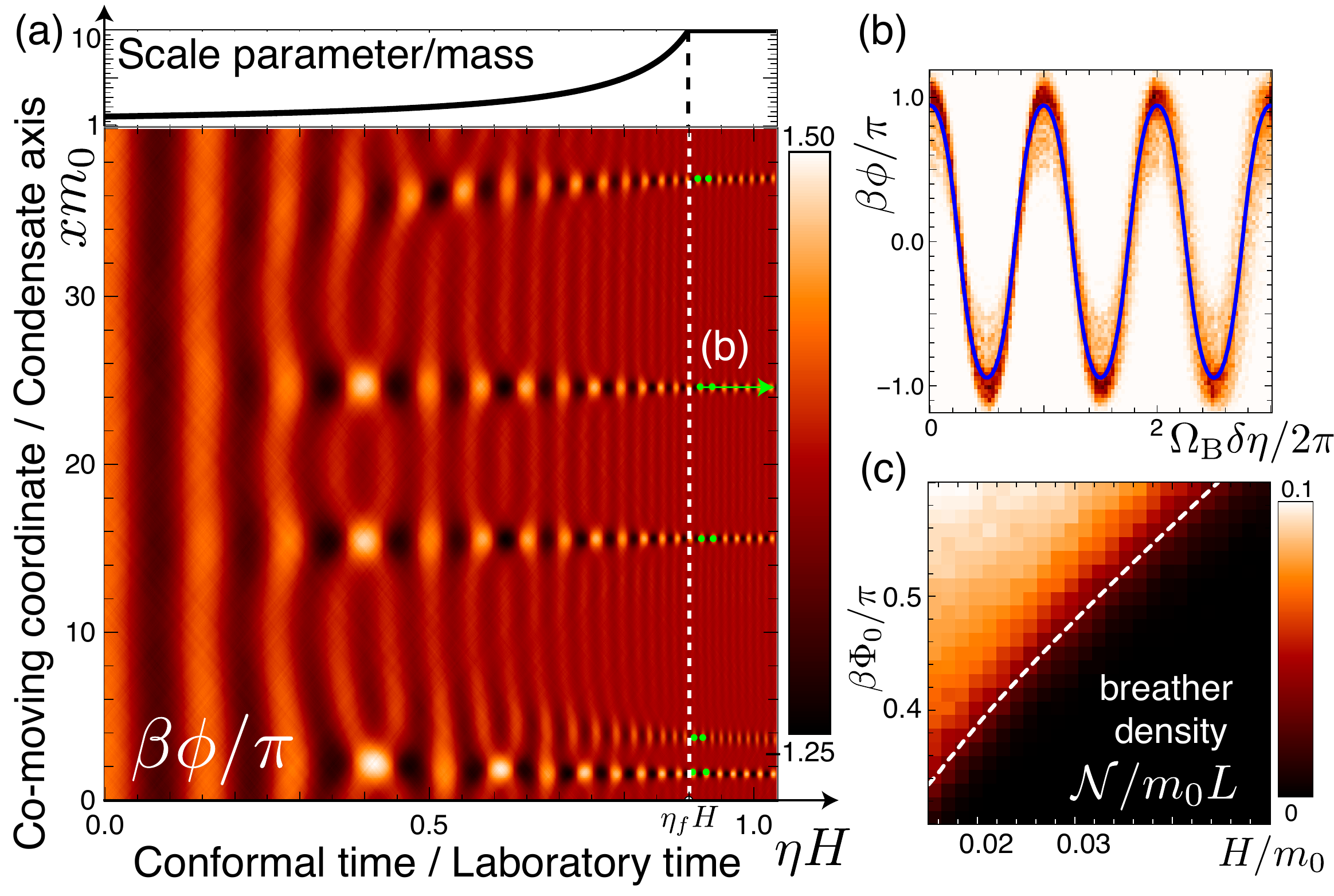}

\caption{(Color online) Emergence of breathers. (a) Plot of a single TWA run
in co-moving coordinates and conformal time ($\beta=0.1$, $\beta\Phi_{0}/\pi=0.65$,
$H/m_{0}=0.02$). One observes the damped Josephson oscillations of
the COM mode and the emergence of breather-like excitations out of
parametrically amplified phonons. As the COM mode is damped away due
to the expansion, standing sine-Gordon breathers are released. These
can be tracked numerically (green dots). (b) Time-evolution of these
breathers tracked at $\delta\eta=0$ denoting the time of tracking
($\eta\geq\eta_{f}$) as a function of $\Omega_{{\rm B}}\delta\eta$
($\Omega_{{\rm B}}$ is the numerically obtained breather frequency).
The spatial center of the breathers is shown. Color code: Distribution
function; Blue, solid line shows the temporal evolution of a standing
breather (corresponding to the mean breather frequency). (c) Density
of created breathers ($\mathcal{N}$ is the number of tracked breathers)
obtained by tracking breathers at $a_{f}=10$. Here, $L=1000$, $m_{0}=0.075$
and $\beta=0.1$. The white, dashed line indicates where $\beta{\rm max}_{\eta}\sqrt{\langle\delta\hat{\phi}^{2}\rangle}=0.2$
(see main text). \label{Fig_Cosmo_Pub_3}}
\end{figure}
While the freezing out of quantum fluctuations is a purely linear
effect, we now discuss a feature, which heavily relies on \emph{interactions}.
From now on, we consider a slow expansion $H/m_{0}\ll1$ and finite
$0<\Phi_{0}\sim\pi/\beta$ (see Fig.$\,$\ref{Fig_Cosmo_Pub_3}a).
At short times $\eta>0$, the global relative phase $\Phi(\eta)$
performs Josephson oscillations (see, e.g., \cite{2012_Neuenhahn}).
However, it is well known that these are parametrically unstable against
small, spatial fluctuations (e.g. \cite{2005_Bochoule,2012_Neuenhahn,1999_Greene}).
Linearizing around $\Phi$, one finds $\hat{\phi}_{k}''+[k^{2}+m_{0}^{2}a^{2}(\eta)\cos\beta\Phi]\hat{\phi}_{k}\simeq0$.
In the static case \cite{2012_Neuenhahn}, this parametric drive leads
to $\hat{\phi}_{k}(\eta)\sim e^{\Gamma(k,\Phi_{0})\eta}$ with $\Gamma\simeq\frac{|k|}{2}\sqrt{\sin^{2}(\beta\Phi_{0}/2)-k^{2}/m^{2}}$
\cite{1999_Greene}(displayed for $\beta\Phi_{0}\lesssim1$). In turn
the fluctuations $\langle\delta\hat{\phi}^{2}\rangle$ grow and at
some point the linearization breaks down. In the semiclassical limit
$\beta\ll1$, it was demonstrated \cite{2012_Neuenhahn} that the
non-linearity of the sine-Gordon equations leads to the formation
of localized patterns in the field $\hat{\phi}(x,\eta)$. These patterns
were identified with 'quasibreather'-solutions of the classical sine-Gordon
equation \cite{2012_Neuenhahn}, which in contrast to usual breathers
have a finite lifetime. The 'quasibreather'-solutions could also explain
the formation of these excitations out of phononic quantum fluctuations. 

Here, for a slow, exponential expansion (in proper time $\tau$),
one observes the formation of similar excitations only for $m_{0}/H$
exceeding a certain, almost sharp threshold (Fig.$\,$\ref{Fig_Cosmo_Pub_3}c,
cf. also \cite{2012_AminPRL}), which depends on $\beta$ and $\Phi_{0}$.
To see this, first note that as long as the linearization of the sine-Gordon
equation applies, all field fluctuations are suppressed as $\hat{\phi}_{k}\sim1/\sqrt{a}$
(including the COM mode). Furthermore, in the course of time, resonant
modes can get shifted out of resonance as a consequence of the expansion.
For a slow expansion, replacing $k\rightarrow k/a$ and $\Phi_{0}\rightarrow\Phi_{0}/\sqrt{a}$
(cf. \cite{1999_Greene,2012_AminPRL}), one finds that $\langle\hat{\phi}_{k}(\eta)\hat{\phi}_{-k}(\eta)\rangle\propto a^{-1}\exp\left[2\int_{0}^{\eta}d\eta'\, a\Gamma(\frac{k}{a},\frac{\Phi_{0}}{\sqrt{a}})\right],$which
is valid, as long as ${\rm Im\Gamma=0}.$ The non-linearity and thus
the formation of localized patterns kicks in only if the fluctuations
$\beta\sqrt{\langle\delta\hat{\phi}^{2}(\eta)\rangle}\simeq[\frac{\beta^{2}}{a(\eta)}\int_{2\pi/L}^{k_{c}}\frac{dk}{2\pi}\frac{\exp[2\int_{0}^{\eta}d\eta'\, a{\rm Re}\Gamma(\frac{k}{a},\frac{\Phi_{0}}{\sqrt{a}})]}{\sqrt{k^{2}+m_{0}^{2}}}]^{1/2}$
exceed a certain value for some $\eta$ (see Fig. $\,$\ref{Fig_Cosmo_Pub_3}c
and \cite{2012_NeuenhahnCosmoAppendix}). Numerically, we find that
this value is of the order $\mathcal{O}(10^{-1})$. 

Close to the creation threshold, once created, these breather-like
excitations persist at the position, where they were 'born' out of
quantum fluctuations. This is in contrast to the 'quasibreathers'
observed in the static case \cite{2012_Neuenhahn} (cf. also \cite{2008_Farhi}).
In the long-time limit, the homogenous part of the field $\Phi$ is
damped away $\propto1/\sqrt{a}$. We find good numerical evidence
that the localized excitations, however, are robust against the expansion
and can be well described as standing (classical) sine-Gordon breathers\emph{
}$\phi_{{\rm B}}$ (see Fig.$\,$\ref{Fig_Cosmo_Pub_3}b). Their typical
distance is set by the maximally amplified wavelength before the non-linearity
sets in, ending the parametric amplification. It seems that at late
times ($a\gg1$), the only effect of the 'adiabatic' expansion ($H/m_{0}\ll1$)
on breathers is a trivial shrinking of the breather period and width
(both $\propto1/m_{0}a(\eta)$) in co-moving coordinates, while their
amplitude stays approximately constant. This can be understood realizing
that the amplitude of a classical sine-Gordon breather $\beta\phi_{{\rm B}}$
is solely determined by the breather parameter $\varphi\in[0,\pi/2]$
(${\rm max}\beta\phi_{{\rm B}}=(2\pi-4\varphi)$, see e.g., \cite{1979_Maki2}).
In the quantum sine-Gordon model, this parameter gets quantized \cite{1975_Dashen},
i.e., it is promoted to a quantum number. However, it is well known
that for a slow change of system parameters (by slow, here, we understand
$\Omega_{{\rm B}}^{-2}\partial\Omega_{{\rm B}}/\partial\eta\ll1$,
where $\Omega_{{\rm B}}=m_{0}a(\eta)\sin\varphi$ is the instantaneous
breather frequency), quantum numbers (and thus the breather amplitude)
are approximatively preserved (cf. \cite{2004_LL1}).
\begin{figure}
\includegraphics[width=1\columnwidth]{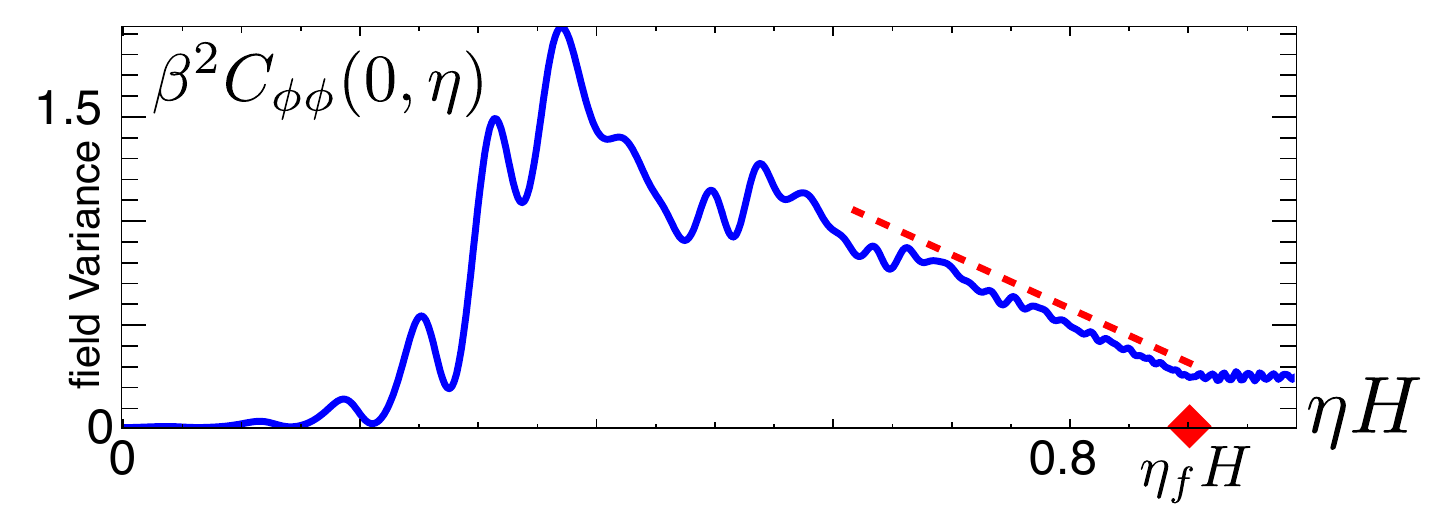}

\caption{(Color online) Plot of the correlation function $C_{\phi\phi}(0,\eta)$.
Clearly, one observes the exponentially growing fluctuations (parametric
resonance). At some point the non-linearity of the sine-Gordon model
leads to the formation of breather-like structures (around the main
peak, cf. Fig.$\,$\ref{Fig_Cosmo_Pub_3}a). While the COM mode is
damped by the expansion, these structures turn into standing sine-Gordon
breathers with a constant amplitude (cf. Fig.$\,$\ref{Fig_Cosmo_Pub_3}b),
which, in physical coordinates simple drift away from each other.
This 'Hubble expansion' is mirrored by the linear decay of the correlation
function as $\propto(1-\eta H)$ at late times. Once the expansion
stops at $\eta_{f}$, the correlation function becomes almost constant.
Here: $\beta=0.1$, $H/m_{0}=0.02$, $m_{0}=0.05$, $L=800$ and $\beta\Phi_{0}=0.7\pi$.\label{Fig_Cosmo_Pub_4}}
\end{figure}
 While for large $a\gg1$ the number and amplitude of breathers remain
constant (per experimental run), they simply move apart from each
other in physical coordinates. This 'Hubble expansion', e.g., can
be observed in the (experimentally accessible) equal time correlation
$C_{\phi\phi}(x,\eta)=\langle\hat{\phi}(x,\eta)\hat{\phi}(0,\eta)\rangle-\langle\hat{\phi}(0,\eta)\rangle^{2}$
(Fig.$\,$\ref{Fig_Cosmo_Pub_4}). Under the assumption that at late
times, the field $\hat{\phi}$ can be described as a set of independent
(standing) sine-Gordon breathers with fixed amplitudes, one obtains
that 
\begin{eqnarray}
C_{\phi\phi}(0,\eta) & \propto & 1-\eta H.\label{eq:-19}
\end{eqnarray}
The linear suppression of $C_{\phi\phi}$ is a direct consequence
of the decreasing breather density in physical coordinates. From a
condensed matter point of view the observation that a suitable protocol
for the tunnel-amplitude prepares a state consisting of independent,
standing sine-Gordon breathers is interesting by itself. While the
analysis here is based on semiclassical considerations (numerically
on the TWA) reliable for $\beta\ll1$, the proposed quantum simulator
could for instance test the stability of classical sine-Gordon breathers
against quantum fluctuations for larger $\beta\sim1$ (cf. \cite{PhysRevD.10.4130}).
Furthermore, a quantum simulation could give insight in the excitation
of 'oscillonic' patterns (as discussed in the cosmology community,
e.g., \cite{2003_Gleiser,2008_Farhi,2010_Hertzberg,2012_AminPRL})
in a scalar quantum field (such as the inflaton field or even the
Higgs field) during a spatial expansion. 

\emph{Conclusions.}\textbf{ }- We demonstrate that tuning the tunnel-amplitude
between a pair of tunnel-coupled 1D condensates, the relative phase
field can simulate an interacting quantum field on an expanding 1+1
space time. The proposed 'quantum simulator' should be realizable
with present cold atom setups. As examples of the quantum many-body
dynamics, which could be investigated, we discussed the freezing out
of phonon modes and the creation of sine-Gordon breathers out of quantum
fluctuations during an exponential FRW-expansion. While the discussion
here is restricted to the semiclassical limit of the underlying quantum
sine-Gordon model, the 'quantum simulator' is meant to explore the
deep quantum regime.

\emph{Acknowledgements}. \textendash{} We thank R. Schützhold for
fruitful discussions related to this work. Financial support by the
Emmy-Noether program is gratefully acknowledged.

\bibliographystyle{apsrev4-1}
\bibliography{/Users/Neuenhahn/Documents/Promotion/BIB/bibPromotion}

\vfill{}

\emph{Supplemental material}. \textendash{} Here, we discuss the dynamics
of massive phonons on the expanding 1+1 dimensional space-time in
conformal time following Eq.$\,$(\ref{EOMModes}). The classical
solution is given by 
\begin{eqnarray}
\phi_{k}(\eta) & = & \frac{C_{k}^{+}}{\sqrt{a}}J_{\nu}(\frac{|k|}{aH})+\frac{C_{k}^{-}}{\sqrt{a}}J_{-\nu}(\frac{|k|}{aH}),\label{eq:-1-2}
\end{eqnarray}
where $\nu=\frac{1}{2}\sqrt{1-4m_{0}^{2}/H^{2}}$ and $J_{\nu}$ denotes
the Bessel function of the first kind. In the following, we restrict
to $H/m_{0}>2$. The quantum mechanical solution follows from Eq.$\,$(\ref{eq:-1-2})
promoting the coefficients $C_{k}^{\pm}$ to operators $\hat{C}_{k}^{\pm}$
and by matching the initial conditions at $\eta=0$ ($a=1$), i.e.,
$\hat{\phi}_{k}(0)=\frac{1}{\sqrt{2\omega_{k}(m_{0})}}(\hat{b}_{k}+\hat{b}_{-k}^{\dagger})$
and $\hat{\phi}_{k}'(0)=-i\sqrt{\frac{\omega_{k}(m_{0})}{2}}(\hat{b}_{k}-\hat{b}_{-k}^{\dagger})$.
Here, $\omega_{k}(m_{0})=\sqrt{k^{2}+m_{0}^{2}}$ is the dispersion
of massive phonons and $\hat{b}_{k}|0\rangle=0$ ($|0\rangle$ denotes
the ground state of massive phonon modes with $k\neq0$ and mass $m_{0}$,
in which the system is initialized). As $J_{\nu}(x>0)\in\mathbb{R}$
for ${\rm Im\nu=0}$, we obtain
\begin{eqnarray}
\hat{C}_{k}^{+} & = & \underbrace{\left[\frac{j_{-\nu}/\sqrt{2\omega_{k}}+iJ_{-\nu}\sqrt{\omega_{k}/2}}{j_{-\nu}J_{\nu}-j_{\nu}J_{-\nu}}\right]}_{\alpha_{k}}\hat{b}_{k}+\alpha_{k}^{\ast}\hat{b}_{-k}^{\dagger},\\
\hat{C}_{k}^{-} & = & \underbrace{\left[\frac{1}{\sqrt{2\omega_{k}}J_{-\nu}}-\alpha_{k}\frac{J_{\nu}}{J_{-\nu}}\right]}_{\beta_{k}}\hat{b}_{k}+\beta_{k}^{\ast}\hat{b}_{-k}^{\dagger},
\end{eqnarray}
where we introduced the abbreviations $j_{\pm\nu}=\left.\partial_{\eta}(a^{-1/2}J_{\pm\nu}\left[\frac{|k|}{aH}\right])\right|_{\eta=0}$
and $J_{\pm\nu}=J_{\pm\nu}\left[\frac{|k|}{H}\right]$. The fluctuation
spectrum yields 
\begin{eqnarray}
\langle0|\hat{\phi}_{k}^{\dagger}(\eta)\hat{\phi}_{k}(\eta)|0\rangle & = & \frac{|\alpha_{k}J_{\nu}\left(\frac{|k|}{aH}\right)+\beta_{k}J_{-\nu}\left(\frac{|k|}{aH}\right)|^{2}}{a(\eta)}.\label{eq:-3-3}
\end{eqnarray}
We are interested in modes starting within the horizon, i.e., with
$|k|\gg H|\nu^{2}-\frac{1}{4}|,m_{0}$. Making use of the limits 
\begin{eqnarray}
\lim_{x/|\nu^{2}-\frac{1}{4}|\rightarrow\infty}J_{\nu}(|x|) & = & -\sqrt{\frac{2}{\pi x}}\sin(\frac{\pi\nu}{2}-x-\frac{\pi}{4}),\nonumber \\
\lim_{x\rightarrow0}J_{\nu}(|x|) & = & \frac{|x|^{\nu}}{2^{\nu}\Gamma(\nu+1)},\label{eq:}
\end{eqnarray}
one obtains for large times $\eta$, such that $|k|/a(\eta)H\ll1$
(and finite $\nu$): 
\begin{eqnarray}
\langle0|\hat{\phi}_{k}^{\dagger}\hat{\phi}_{k}|0\rangle & \simeq & \frac{1}{H^{1-2\nu}a(\eta)^{1-2\nu}}\frac{1}{|k|^{2\nu}}.\label{eq:-1}
\end{eqnarray}

\end{document}